\documentclass[aps,nofootinbib,11pt,onecolumn]{revtex4-1}

\usepackage{epsfig}
\usepackage{amsmath,amssymb,amsfonts}
\usepackage{color}

\begin{document}
\title{Radiation damping in pulsed Gaussian beams}
\author{Chris Harvey}
\email{christopher.harvey@physics.umu.se}
\author{Mattias Marklund}
\affiliation{Department of Physics, Ume\aa\ University, SE-90187 Ume\aa, Sweden}

\begin{abstract}
We consider the effects of radiation damping on the electron dynamics in a Gaussian beam model of a laser field.  For high intensities, i.e. with dimensionless intensity $a_0 \gg 1$, it is found that the dynamics divide into three regimes.  For low energy electrons (low initial $\gamma$-factor, $\gamma_0$) the radiation damping effects are negligible.  At higher energies, but still at $2\gamma_0< a_0$, the damping alters the final displacement and the net energy change of the electron.  For $2 \gamma_0 >a_0$ one is in a regime of radiation reaction induced electron capture.  This capture is found to be stable with respect to the spatial properties of the electron beam and results in a significant energy loss of the electrons.  In this regime the plane wave model of the laser field provides a good description of the dynamics, whereas for lower energies the Gaussian beam and plane wave models differ significantly.  Finally the dynamics are considered for the case of an XFEL field.  It is found that the significantly lower intensities of such fields inhibits the damping effects.
\end{abstract}
\maketitle
\section{Introduction}
The past few years have seen many advances in laser technology resulting in lasers of unprecedented powers and intensities, the current record being approximately $10^{22}$W/cm$^2$ \cite{Yanovsky:2008}.  This trend is expected to continue throughout the next few years, such as with the European Extreme Light Infrastructure (ELI), a facility that may deliver intensities as high as $10^{26}$W/cm$^2$ \cite{ELI}.  Such extreme intensities will allow the probing of fundamental physics in previously inaccessible regimes.

In fields of such high intensities, an electron will be accelerated so strongly that its own radiation emission may significantly affect its motion.  This opens up the possibility of testing experimentally the classical theory of radiation reaction in such a context.  Moreover, even when the classical theory is no longer directly relevant, such as when considering laser induced strong field QED processes, an understanding of radiation damping effects is still important.  This is because many such QED processes either have close classical analogues (e.g. nonlinear Compton scattering \cite{Harvey:2009ry, Heinzl:2009nd}) or the current models describing them incorporate classical theory.  An example of the latter is the topical area of runaway QED cascading in strong laser pulses.  The numerical modelling of this phenomena often involves propagating the particles classically through the laser, whilst altering the particle number at each time step via quantum transition rates (see e.g. \cite{Elkina:2010up, Nerush:2011xr}).

An important factor that needs to be taken into consideration is the description we choose for the laser pulse.  Many recent discussions of radiation damping effects have restricted their consideration to plane wave models of the laser field \cite{Koga:2004, Hadad:2010mt, Harvey:2010ns}.  However, the new generation of high intensity facilities will, in part, achieve their high outputs by a strong focussing of the laser pulse.  This suggests that the spatial properties of the beam will become more important and so we should adopt a more realistic description of the field.  In this paper will will use a Gaussian paraxial beam approximation, including terms up to fifth order in the expansion parameter.  This beam model describes the spatial as well as the temporal properties of the beam, giving us an accurate description of the field profile.  We will also compare our results with those of the plane wave model, which will allow us to evaluate the accuracy of such a description.

In this paper we have two aims.  Firstly, to assess the significance of radiation damping on the electron dynamics in a realistic beam profile.  Secondly, to look for regimes where the radiation reaction effects are particularly prominent, allowing for an easy experiment testing of the classical theory.

\section{Theory}
\subsection{Equation of motion}
A charged particle undergoing acceleration (e.g. due to the presence of a laser field) will emit electromagnetic radiation.  The emission of this radiation will result in a momentum loss for the particle, thus subsequently affecting its motion.  The classical action describing such a system is \cite{Jackson:1999,Landau:1987}
\begin{eqnarray} 
  S = -m \int d\tau - e \int d^4 x \, j^\mu A_\mu - \frac{1}{4}
  \int d^4 \, x F_{\mu\nu} F^{\mu\nu},\label{action}\end{eqnarray}
where $m$ is the mass (from hereon taken to be the electron mass), $e$ the charge, $j^\mu$ the four-current and we have adopted units where the speed of light is unity, $c=1$.  The gauge potential and electromagnetic field tensor are denoted by $A^\mu$ and $F^{\mu\nu}$ respectively.  Varying (\ref{action}) with respect to $A^\mu$ and $x^\mu$ gives us the governing equations
\begin{eqnarray}
\partial_\mu F^{\mu\nu} &=& j^\nu,\\
m\dot{u}^\mu &=& e F^{\mu\nu}u_\nu \equiv F^\mu.
\end{eqnarray}
In the second equation $u^\mu$ is the particle four-velocity and the dot denotes differentiation with respect to proper time $\tau$.  The electromagnetic tensor $F^{\mu\nu}$ may be decomposed into the sum of the external field (in this case the laser field) $F^{\mu\nu}_{\textrm{ext}}$ and the radiation field emitted by the electron $F^{\mu\nu}_{\textrm{rad}}$
\begin{eqnarray}
F^{\mu\nu}=F^{\mu\nu}_{\textrm{ext}}+F^{\mu\nu}_{\textrm{rad}}.
\end{eqnarray}
The resulting equation of motion, after calculating $F^{\mu\nu}_{\textrm{rad}}$, is the Lorentz Abraham Dirac (LAD) equation \cite{Lorentz:1905,Abraham:1905,Dirac:1938nz}
\begin{eqnarray}
  m \dot{u}^\mu =  e F_{\mathrm{ext}}^{\mu\nu} u_\nu  -  \frac{2}{3}
  \frac{e^2}{4\pi} (u^\mu \ddot{u}^\nu - u^\nu \ddot{u}^\mu) \,
  u_\nu.\label{lad}
\end{eqnarray}
This equation is infamous due to the presence of the $\ddot{u}$ terms on the right hand side which give rise to (unphysical) runaway solutions.  A common way to remove this problem is to approximate $\ddot{u}$ using the first term in (\ref{lad}) (i.e. the Lorentz force expression).  This results in the Landau Lifshitz (LL) equation \cite{Landau:1987}
\begin{eqnarray}
 \dot{u}^\mu = \frac{e}{m} F^{\mu\nu} u_\nu + r_0 \left\{
  \frac{e}{m^2} \dot{F}^{\mu\nu} u_\nu + \frac{e^2}{m^3}
  F^{\mu\alpha}F_\alpha^{\;\;\nu} u_\nu - \frac{e^2}{m^3} u_\alpha
  F^{\alpha\nu}F_\nu^{\;\;\beta} u_\beta \, u^\mu \right\},\label{LL1}
\end{eqnarray}
which is a perturbative expansion of LAD to first order in the coupling $r_0\equiv 2\alpha/3$, where $\alpha=e^2/4\pi$ is the fine structure constant, and has the benefit of no longer exhibiting the runaway solutions.  

For a laser beam described by the wave vector $k^\mu$ we define a dimensionless measure of laser intensity (see \cite{Heinzl:2008rh})
\begin{eqnarray}
a_0^2 \equiv \frac{e^2}{m^2}\frac{\langle F^{\mu}_{\phantom{\mu}\nu}p^{\nu}_0\rangle^2}{(k\cdot p_0)^2},\label{a0}
\end{eqnarray}
where $p_0^\mu=mu_0^{\mu}$ is the initial electron four-momentum and the brackets $\langle \hdots\rangle$ denote the maximum value of the enclosed quantity over a laser cycle.  For a plane wave laser in the lab frame, this takes the more familiar form
\begin{eqnarray}
a_0=\frac{eA_0}{\omega m },
\end{eqnarray}
where $\omega$ is the laser frequency and $A_0$ the magnitude of the electrical field strength.  

To neglect the influence of radiation damping we set $r_0=0$.  Then the equation of motion (\ref{LL1}) takes the form
\begin{eqnarray}
m\dot{u}^\mu=eF^{\mu\nu}u_\nu, \label{LF}
\end{eqnarray}
which is, of course, the Lorentz force equation.  Using the solution to (\ref{LF}), the standard measure of radiated energy loss is given by Larmor's formula\footnote{Using just the Lorentz force solution for $u$ is consistent with the LL equation being an expansion truncated to first order in $r_0$ \cite{Harvey:2011}.}
\begin{eqnarray}
P=r_0\dot{u}^2,
\end{eqnarray}
where $P$ is the radiated power.  The energy loss of the electron over a laser cycle is given by
\begin{eqnarray}
R\equiv\frac{P}{\omega m}= r_0 \frac{\omega}{m}a_0^2 \gamma (1+\beta ). \label{R}
\end{eqnarray}
When this parameter reaches unity we are in the ``radiation dominated regime'' \cite{DiPiazza:2009zz} where the radiation damping effects are of the same magnitude as the Lorentz force.  In order for the LL equation to be valid as a perturbative expansion of LAD we require that the radiation damping term is much smaller than the Lorentz force term.  Since it is the final term of (\ref{LL1}) that gives us the biggest contribution to the damping, we require that
\begin{eqnarray}
r_0 e^2 u_\alpha F^{\alpha\nu}F_\nu^{\;\;\beta} u_\beta \, u^\mu /m^3
\ll e F^{\mu\nu}u_\nu /m,
\end{eqnarray}
which, for a head-on collision in the lab frame between the electron and a plane wave laser field, gives us the constraint
\begin{eqnarray}
r_0 \omega^2 a_0^2 \gamma^3/m \ll \omega a_0\gamma.
\end{eqnarray}
For an optical laser this becomes
\begin{eqnarray}
a_0\gamma^2\ll 10^8 \label{llconstraint}.
\end{eqnarray}
Indeed, in \cite{Hadad:2010mt} the authors find that the solutions to the LAD and LL expressions begin to differ even at an order of magnitude below this estimate.

As well as ensuring that we stay in a regime where the derivation of the LL equation is valid, we must also take a moment to ensure that we keep within the domain of classical physics.  A common measure of the importance of quantum effects is to consider the work done by the laser field over the distance of a Compton wavelength
\begin{eqnarray}
\chi\equiv \frac{e \sqrt{(F^{\mu\nu}p_\nu)^2}}{m^3}.
\end{eqnarray}
This was first introduced in \cite{Nikishov:1964zza, Narozhnyi:1965} and is expressed here in units where $\hbar =1$.  Quantum effects become important when this measure approaches unity.  In the lab frame we find that to be in the classical regime we require 
\begin{eqnarray}
a_0\gamma \omega \ll m,\label{qedconstraint}
\end{eqnarray}
which, for an optical laser, can be a more stringent requirement than (\ref{llconstraint}).  

\subsection{Description of the field}
We describe our laser field using a pulsed Gaussian beam model, in a similar way to Salamin \textit{et al} in \cite{Salamin:2002dd}.  Proceeding in the same way as McDonald \cite{McDonaldsnotes} we begin by adopting the Lorentz gauge
\begin{eqnarray}
\frac{\partial\phi}{\partial t}+\boldsymbol{\nabla}\cdot\boldsymbol{A}=0,\label{lorentzgauge}
\end{eqnarray}
and dictate that any vector potential describing our laser field must satisfy the vacuum wave equation
\begin{eqnarray}
\nabla^2\boldsymbol{A}=\frac{\partial^2\boldsymbol{A}}{\partial t^2}. \label{vacuumwaveeqn}
\end{eqnarray}
We take the laser to propagate in the $+z$ direction and assume a generic potential, linearly polarised in $x$
\begin{eqnarray}
\boldsymbol{A}=\boldsymbol{\hat{x}}A_0 g(\eta)\psi (x,y,z)e^{-ikz},\label{A1}
\end{eqnarray}
where $A_0$ is the wave amplitude, $\eta =\omega t-kz$, and $g$ is a generic pulse shape function.  Inserting (\ref{A1}) into (\ref{vacuumwaveeqn}) gives us
\begin{eqnarray}
\nabla^2\psi-2ik\frac{\partial\psi}{\partial z} 
\left ( 1-i\frac{g^\prime}{g}\right )=0, \label{psieqn}
\end{eqnarray}
where $g^\prime =dg/d\eta$.  In general it is hard to satisfy (\ref{psieqn}) since $\psi$ is a function of $(x,y,z)$ and $g$ is a function of the phase $\eta$.  To proceed we begin by rescaling our coordinates
\begin{eqnarray}
\xi\equiv\frac{x}{w_0}, \quad \nu\equiv\frac{y}{w_0}, \quad \zeta\equiv\frac{z}{z_r},
\end{eqnarray}
making them dimensionless.  Here $w_0$ is the beam waist diameter and $z_r=kw_0^2/2$ is the Rayleigh length.  Following \cite{McDonaldsnotes} we specify that the pulse shape function satisfies
\begin{eqnarray}
g^\prime \ll g. \label{gconstraint}
\end{eqnarray}
Equation (\ref{psieqn}) can then be approximated by
\begin{eqnarray}
\nabla_\perp^2\psi -4i\frac{\partial\psi}{\partial\zeta}+\theta_0^2
\frac{\partial^2\psi}{\partial\zeta^2}=0, \label{psieqnapprox}
\end{eqnarray}
where
\begin{eqnarray}
\nabla_\perp^2 =\frac{\partial^2}{\partial\xi^2}+\frac{\partial^2}{\partial\nu^2},\quad
\psi=\psi(\xi, \nu, \zeta ),
\end{eqnarray}
and we have introduced the aspect ratio $\theta_0 =w_0/z_r$ which, when small, closely approximates the beam diffraction angle.  Since $\theta_0$ is typically small, we can expand $\psi$ in the series
\begin{eqnarray}
\psi=\psi_0+\theta_0^2\psi_2+\theta_0^4\psi_4+\ldots.
\end{eqnarray}
Equating coefficients of $\theta_0$ we have, from (\ref{psieqnapprox}),
\begin{eqnarray}
\nabla_\perp^2\psi_0-4i\frac{\partial \psi_0}{\partial\zeta} &=& 0, \label{E1}\\
\nabla_\perp^2\psi_2-4i\frac{\partial \psi_2}{\partial\zeta}+\frac{\partial^2 \psi_0}{\partial\zeta^2} &=& 0,\label{E2}\\
\nabla_\perp^2\psi_4-4i\frac{\partial \psi_4}{\partial\zeta}+\frac{\partial^2 \psi_2}{\partial\zeta^2} &=& 0,\label{E3}\\
\textrm{etc.}\nonumber
\end{eqnarray}
Equation (\ref{E1}) is the paraxial wave equation from traditional Gaussian beam theory.  Its solution is the well known first order Gaussian beam solution
\begin{eqnarray}
\psi_0 =fe^{-f\rho^2},
\end{eqnarray}
where
\begin{eqnarray}
f=\frac{1}{\sqrt{1+\zeta^2}}e^{i\arctan\zeta},\quad \rho^2=\xi^2+\nu^2.
\end{eqnarray}
The solution to (\ref{E2}) was originally found by Davis \cite{Davis:1979}
\begin{eqnarray}
\psi_2=\left ( \frac{f}{2}+\frac{f^3\rho^4}{4}\right )\psi_0,
\end{eqnarray}
and Barton and Alexander \cite{Barton:1989} proceeded to find the solution to (\ref{E3})
\begin{eqnarray}
\psi_4=\frac{1}{32}(12f^2-6f^4\rho^4-4f^5\rho^6+f^6\rho^8)\psi_0.
\end{eqnarray}
Analogously to the case of the vector potential (\ref{A1}), we assume that the scalar potential can be written in the form
\begin{eqnarray}
\phi(t,x,y,z)=g(\eta)\Phi(x,y,z)e^{i\eta}.
\end{eqnarray}
Then the Lorentz gauge condition (\ref{lorentzgauge}) gives us
\begin{eqnarray}
\frac{\partial\phi}{\partial t}=i\omega\phi\left (1-i\frac{g^\prime}{g}\right )\approx i\omega\phi,
\end{eqnarray}
which means that
\begin{eqnarray}
\phi=\frac{i}{k}\boldsymbol{\nabla}\cdot\boldsymbol{A}.
\end{eqnarray}
Thus our electric and magnetic field components may be found from (\ref{A1}) via
\begin{eqnarray}
\boldsymbol{E} &=& -ik\boldsymbol{A}-\frac{i}{k}\boldsymbol{\nabla}(\boldsymbol{\nabla}\cdot\boldsymbol{A}),\\
\boldsymbol{B} &=& \boldsymbol{\nabla}\times\boldsymbol{A},
\end{eqnarray}
(for details of the calculation see \cite{Salamin:2002dd, Barton:1989}).  Taking the real part of the resulting expressions gives us (to fifth order in $\theta_0$)
\begin{eqnarray}
E_x &=& P\bigg(S_0 + \frac{\theta_0^2}{4}\left [4\xi^2 S_2 -\rho^4S_3\right ] \nonumber\\
&&\quad+\frac{\theta_0^4}{32}\left [  4S_2-8\rho^2 S_3-2\rho^2(\rho^2-16\xi^2)S_4 -4\rho^4(\rho^2+2\xi^2)S_5 +\rho^8S_6\right ]\bigg),\label{fieldEx}\\
E_y &=& P\xi\nu\bigg(\theta_0^2 S_2+\frac{\theta_0^4}{4}\left [ 4\rho^2 S_4-\rho^4 S_5 \right ]         \bigg),\label{fieldEy}\\
E_z &=& P\xi\bigg( \theta_0 C_1  +\frac{\theta_0^3}{4}\left [ -2C_2+4\rho^2 C_3-\rho^4 C_4 \right ] \nonumber \\
&&\quad+\frac{\theta_0^5}{32} \left [ -12C_3 -12\rho^2 C_4 +34\rho^4 C_5 -12\rho^6 C_6 +\rho^8 C_7   \right ]  \bigg),\label{fieldEz}\\
B_x &=& 0,\label{fieldBx}\\
B_y &=& P\bigg( S_0+ \frac{\theta_0^2}{4}\left [  2\rho^2 S_2-\rho^4S_3\right ]
+\frac{\theta_0^4}{32}\left [  - 4S_2+8\rho^2S_3+10\rho^4 S_4-8\rho^6 S_5 +\rho^8 S_6\right ]   \bigg),\label{fieldBy}\\
B_z &=& P\nu\bigg( \theta_0 C_1 +\frac{\theta_0^3}{4}\left [ 2C_2+2\rho^2 C_3-\rho^4 C_4  \right ]  \nonumber\\ 
&&\quad+\frac{\theta_0^5}{32}\left [  12C_3 +12\rho^2 C_4 +6\rho^4C_5 -8\rho^6 C_6 +\rho^8C_7   \right ]\bigg),\label{fieldBz}
\end{eqnarray}
where the prefactor is given by
\begin{eqnarray}
P=A_0 \frac{w_0}{w}g(\eta )\textrm{exp}\big( -\frac{r^2}{w^2} \big), \quad r^2=x^2+y^2.
\end{eqnarray}
Here $w=w(z)$ is a measure of the beam diameter according to
\begin{eqnarray}
w(z)=w_0\sqrt{1+\bigg(\frac{z}{z_r}\bigg)^2},
\end{eqnarray}
and the functions $S_j$ and $C_j$ are defined
\begin{eqnarray}
S_j &=& \bigg( \frac{w_0}{w} \bigg)^j \sin\Theta, \\
C_j &=& \bigg( \frac{w_0}{w} \bigg)^j \cos\Theta, 
\end{eqnarray}
where
\begin{eqnarray}
\Theta =\eta-\frac{kr^2}{2H}+(j+1)\arctan\zeta,
\end{eqnarray}
where $H=z+z_r^2/z$ is the radius of curvature of the field.

\begin{figure}[!h]
\includegraphics[scale=1.0,clip=true,viewport=90 230 600 580]{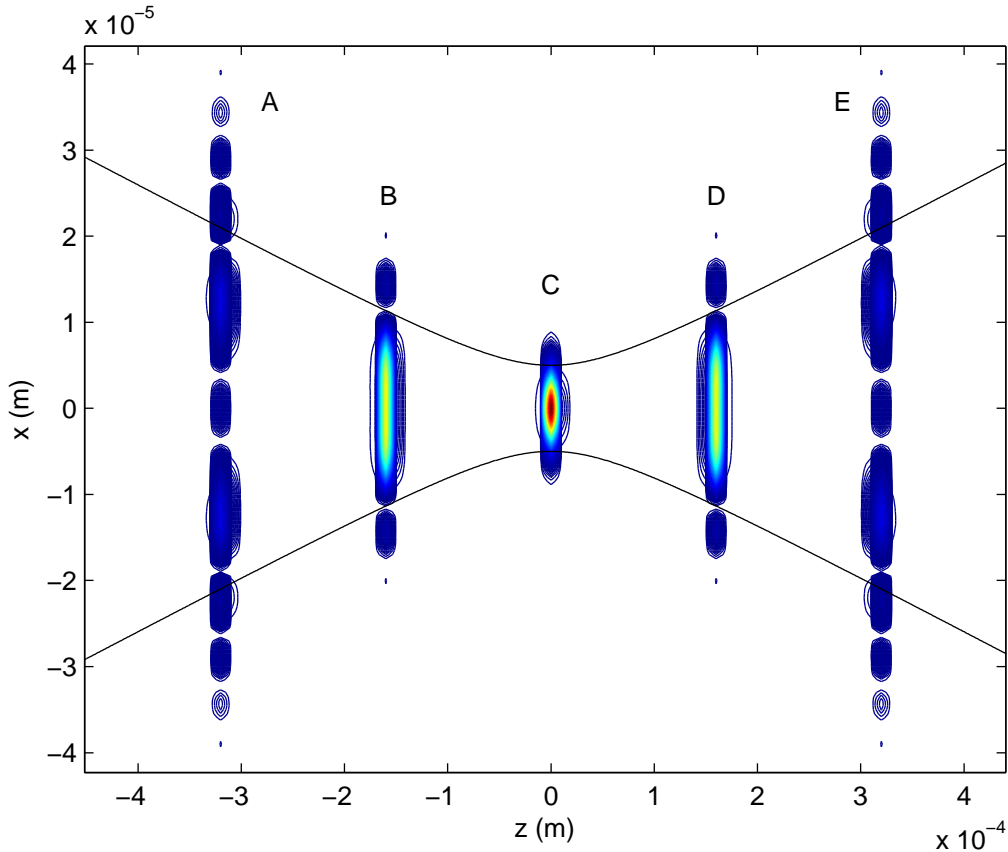}
\caption{\label{fig:pulse} Plot showing the energy density ($(E^2+B^2)/2$) at different times. A: $\omega t=-2000$, B: $\omega t=-1000$, C: $\omega t=+0.5$, D: $\omega t=+1000$, E: $\omega t=+2000$.  Parameters are $\lambda=1\mu m$, $w_0=5\mu m$, $\eta_0 =20$.  The solid black lines show the boundary of the beam waist $w_0$.
}
\end{figure}

It has been shown in \cite{Salamin:2002dd} that, to accurately describe the properties of an optical beam for realistic parameter values ($w_0\sim 5\mu$m), it is necessary to keep  terms up to at least $\mathcal{O}(\theta_0^3)$.  With this in mind we will henceforth work with the full expressions (to order $\theta_0^5$) given above.  (However, in the case of an X-ray free electron laser (XFEL) the beam is much less tightly focused, typically having a waist that is many wavelengths in diameter.  In such cases the expansion parameter, $\theta_0$, is small enough that one only needs to use the first term without corrections.)  All that remains is to define our pulse shape function $g(\eta )$.  A common choice when studying plane wave dynamics is to use a Gaussian pulse profile $g=\textrm{exp}[-(\eta/\eta_0)^2]$, where $\eta_0$ is the number of cycles in the pulse.  However, this function does not always satisfy the constraint (\ref{gconstraint}).  Instead, in this study we adopt the profile recommended by McDonald \cite{McDonaldsnotes}
\begin{eqnarray}
g(\eta) =\textrm{sech}\bigg(\frac{\eta}{\eta_0} \bigg).
\end{eqnarray}
In Figure \ref{fig:pulse} we show the laser pulse intensity for typical parameter values and at various times. 

\subsection{The set-up}
Our aim is both to assess the significance of radiation damping in different regimes and to focus on those regimes where the effects are most prominent.  The latter cases will allow for experimental tests of the classical theory.  We see from (\ref{LL1}) that the coupling $r_0$ between the radiation damping force and the Lorentz force is very small.  In order to compensate for this we need either for the damping force to be large, or for the electron to experience the damping force for a long time period. For the damping force to be large the electron must undergo a large acceleration.  In practice this means that the electron must see a laser field of high intensity -- either by having a large $a_0$ or by having a head on (or near head on) collision between a moderately high intensity laser and a moderately high energy electron (moderately high meaning that we adhere to the physical constraints (\ref{llconstraint},\ref{qedconstraint})).  The other option is for the electron to spend a longer time period in the laser field.  This is best achieved via a capture and acceleration scenario (CAS) \cite{Wang:1998fb, Wang:1999yea, Wang:2002zza} where the electron enters the field and gets captured near the focus and subsequently accelerated.  For such a situation to occur we should have a high energy electron interacting with the laser pulse in a same direction (or near same direction) collision.  Although in the electron's frame the laser is not as strong as it would be for a head on collision, the fact that the electron is fast moving means that it will travel along with the laser pulse for a significant period of time and so, even though they are smaller, the damping effects will accumulate.  The initial conditions must be chosen such that the electron is captured by the beam, rather than being ejected due to the ponderomotive force.  This is generally achieved by having a suitable balance between $a_0$ and the initial electron energy.  It is also advantageous to have our pulse length $\eta_0$ as large as possible (but, of course, the pulse length is constrained by the total amount of energy the laser facilities are able to deliver).  

\begin{figure}[!h]
\includegraphics[scale=1.0,clip=true,viewport=90 230 600 580]{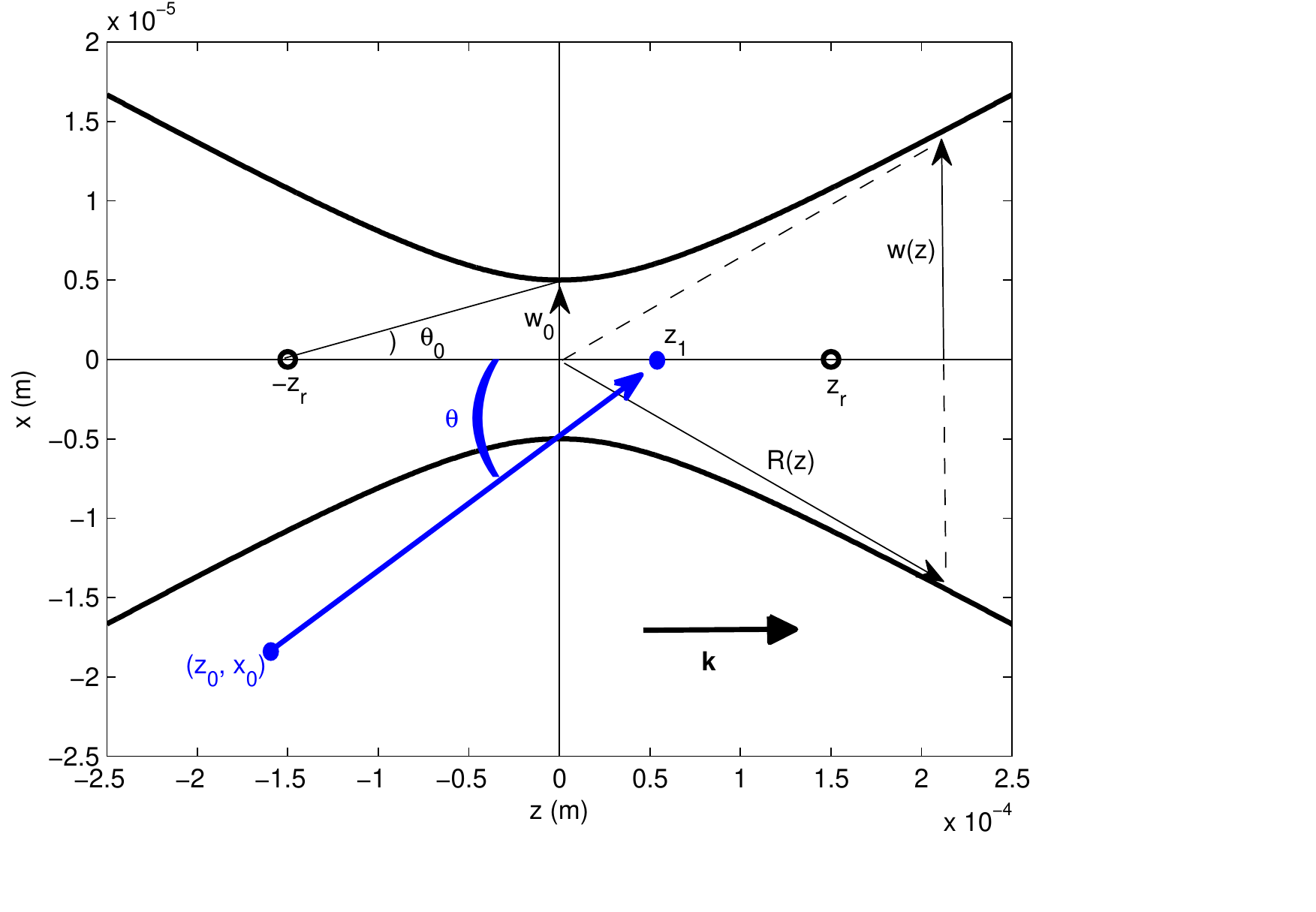}
\caption{\label{fig:geometry} Schematic of our simulation set up shown in the $y=0$ plane.  The laser is propagating in the $\boldsymbol{k}=+z$ direction and the electron starts at $(z_0, x_0)$ heading towards $(z_1, 0)$ at an angle $\theta$ to the $z$ axis.}
\end{figure}

In Figure \ref{fig:geometry} we give a schematic of our set up.  The simulation will take place in three dimensions, but for simplicity we will consider things here in the $y=0$ plane.  We start our simulation when the electron is at the point $(z_0, x_0)$ with energy $\gamma_0$ moving towards the point $(z_1, 0)$, making an angle $\theta$ with the beam axis.  When $z_1=0$ the electron is aimed at the centre of the pulse, otherwise ($z_1\ne 0$) the electron is sent towards the right or left of the focus.  The beam is propagating in the $+z$ direction; therefore collisions where $\theta\sim 0$ will be referred to as `near same direction' and cases where $\theta\sim 180$ `near head on'.

We will solve the equation of motion (\ref{LL1}) numerically using a second order leapfrog method.  To make the process less computationally expensive, we exclude the first of the radiative correction terms (the term proportional to $\dot{F}^{\mu\nu}$) from our calculation.  This term goes like $r_0 a_0 \gamma$ which, for an optical laser, even at the maximally allowed value of $a_0 \gamma$ (\ref{qedconstraint}) is still 2 orders of magnitude smaller than the Lorentz force term.  Test runs of our code indicate that this is a justified approximation.

\section{Results}
\subsection{Plane wave dynamics}
Before considering the full Gaussian beam model, we will first devote our attention to a simpler case -- that of the plane wave.  We obtain plane wave fields in the limit where the beam waist becomes large, $w_0\to\infty$.  Then the laser field takes the configuration
\begin{eqnarray}
E_x &=& A_0 g(\eta)\sin\eta,\quad E_y =0,\quad E_z=0,\\
B_x &=&  0, B_y =A_0 g(\eta)\sin\eta,\quad\quad B_z=0.
\end{eqnarray}
Plane wave models are commonly used when modelling laser-electron interactions since they are simple enough to allow the properties of many relevant physical phenomena to be calculated analytically, whilst still retaining important characteristics of the laser field (such as time-dependence).  From our perspective, an additional reason for considering plane waves is that the analytical solution to the LL equation is known \cite{DiPiazza:2008}, thus allowing us to benchmark our codes.  However, when working with plane wave models one must proceed with caution.  Plane waves are infinite in their spatial extent and therefore unsuited to situations where the laser-electron interaction occurs over a long spatial distance.  For example, a high energy electron propagating in the same direction as the laser will spend a long time in the laser pulse since it will be travelling at nearly the speed of light.  In such circumstances a long interaction time equates to a long spatial \textit{distance} of interaction in the propagation direction.  It can be seen from (\ref{fieldEx}-\ref{fieldBz}) that, even over a modest distance in the $z$-direction, a realistic laser field will become significantly damped (decaying like $1/w\sim 1-z^2/z_r^2$), whereas the plane wave model will maintain a constant peak amplitude throughout.  Because of this we will only consider the case of a near head on collision in our plane wave analysis.

\begin{figure}[!h]
\includegraphics[scale=0.8,clip=true,viewport=00 250 600 580]{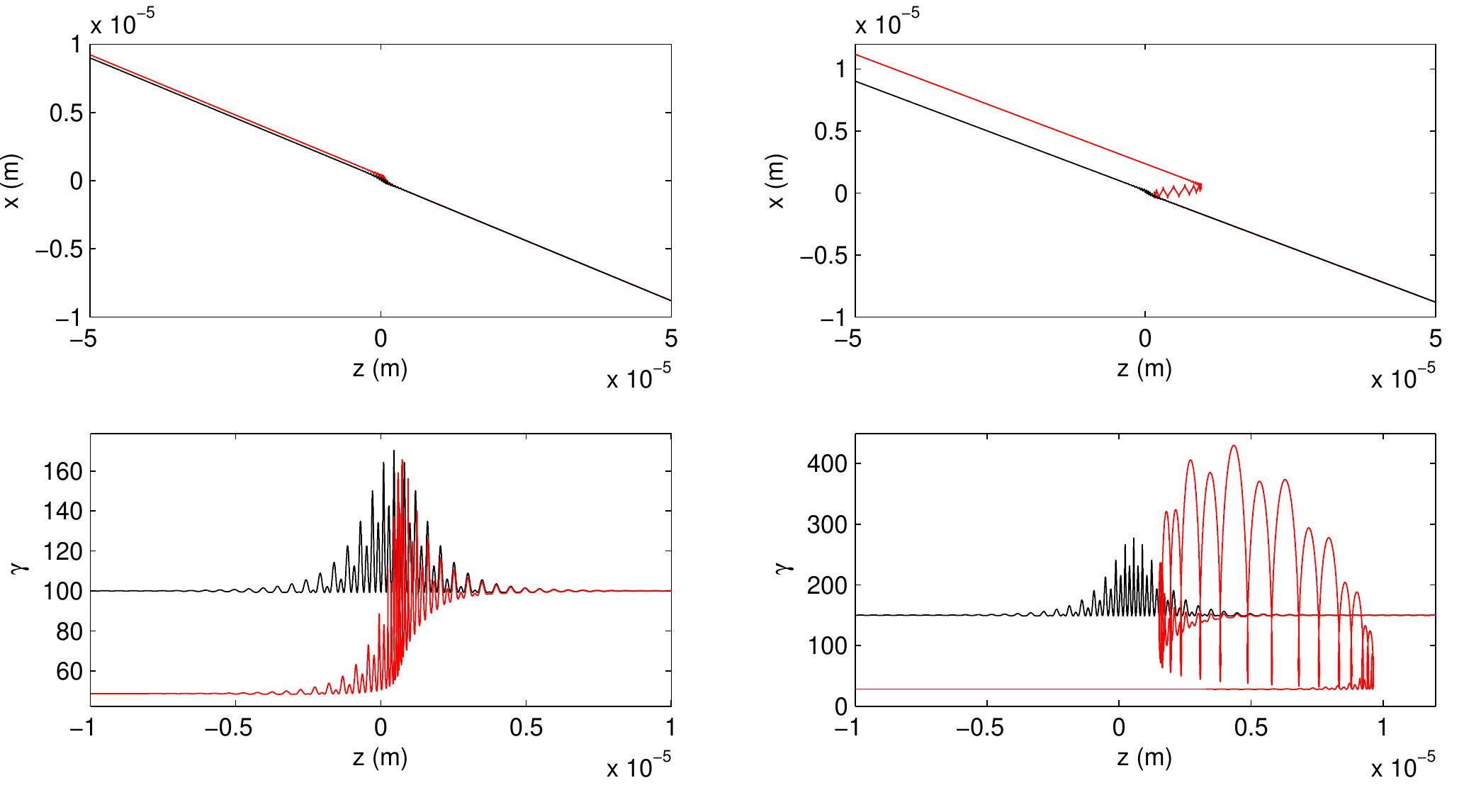}
\caption{\label{fig:FigurePW} Near head on collision ($\theta=170$ degrees) for plane wave laser field.  Left: $a_0=150$, $\gamma_0 =100$, right:  $a_0=250$, $\gamma_0 =150$; $\eta_0 =20$ for both cases.  Top plots show trajectories and bottom plots show corresponding electron energies ($\gamma$-factors).  Black lines: undamped solutions (Lorentz force), red lines: damped solutions (LL).}
\end{figure}

\begin{figure}[ht!]
\includegraphics[scale=0.8,clip=true,viewport=10 215 600 650]{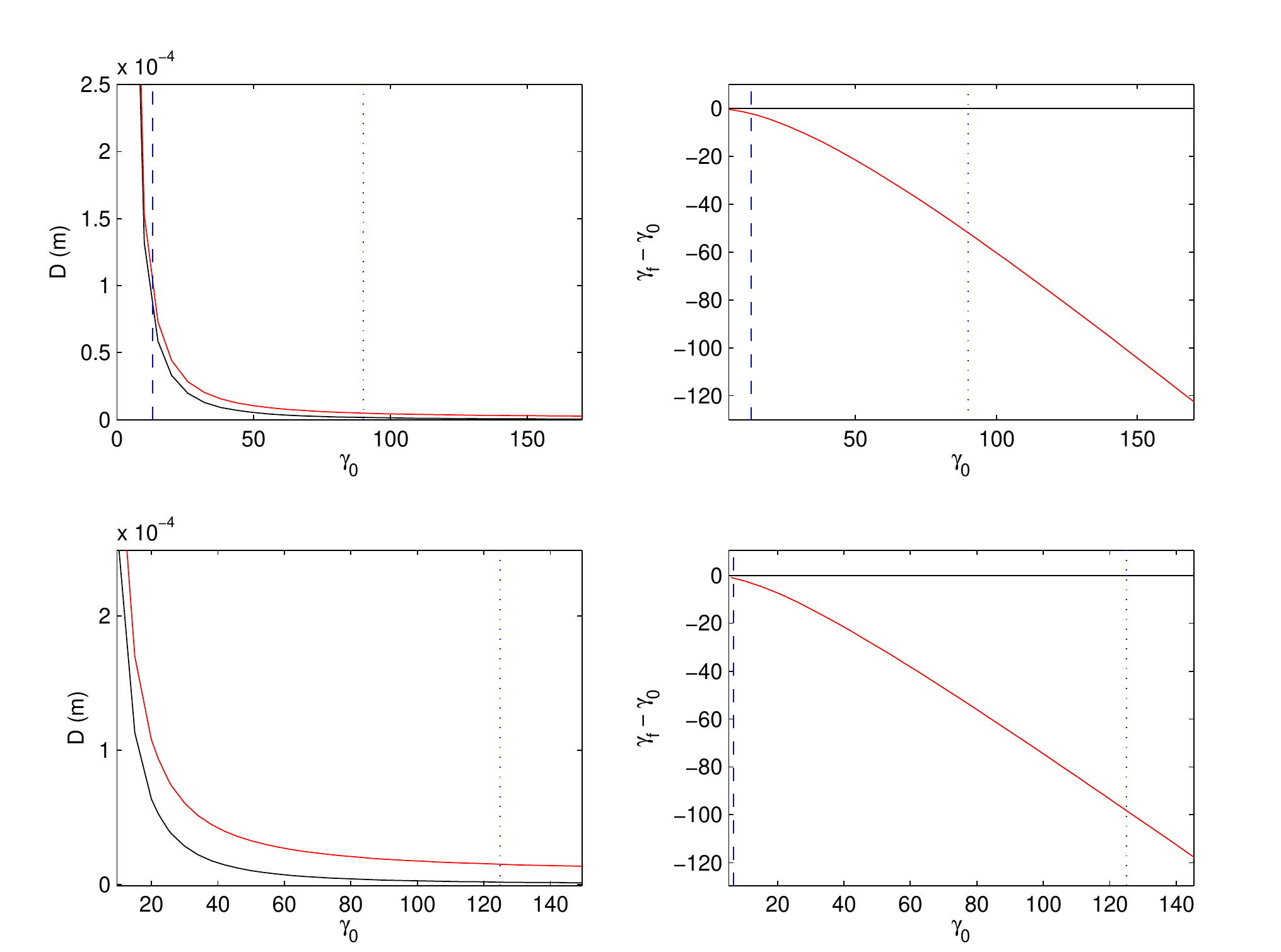}
\caption{\label{fig:pw_disp} Left plots: longitudinal displacement $D$ of the electron as a function of $\gamma_0$ for a plane wave, near same direction collision ($\theta=170$ degrees).  Right plots: net energy change $\gamma_f-\gamma_0$ as a function of $\gamma_0$. Top plots have parameters $a_0=180$, $\eta_0 =20$ and bottom plots $a_0=250$, $\eta_0 =20$.  Black lines:  undamped solutions, red lines: damped solutions, blue dotted lines indicate $1\%$ of $R$, blue dashed lines: $2\gamma_0=a_0$.
}
\end{figure}

The left hand plots of Figure \ref{fig:FigurePW} show the case of a near head on collision ($\theta =170^\circ$) between  an electron of initial energy $\gamma_0=100$ and a plane wave of intensity $a_0=150$.  The pulse parameter is chosen such that the laser has 20 cycles ($\eta_0 =20$) which, for the 1$\mu$m wavelength we are considering, corresponds to a pulse duration of approximately 10fs.  We can see that for these parameter values radiation damping only has a minimal effect on the electron's trajectory.  Nevertheless, as can be seen from the lower left plot, the electron loses approximately 50\% of its energy due to radiation damping.  This is consistent with the analysis presented by Koga \textit{et al} \cite{Koga:2005} (see also \cite{Harvey:2010ns}).  The right hand plots of Figure \ref{fig:FigurePW} show the same quantities but for different parameter values -- this time $a_0=250$, $\gamma_0=150$.  These parameters are towards the higher end of the range allowed by our conditions (\ref{llconstraint}, \ref{qedconstraint}).  What is most noticeable in these plots is that the radiation damping is now strong enough to reflect the electron so that for a while it co-propagates with the laser.  The onset of this reflection regime was first identified by Di Piazza \textit{et al} \cite{DiPiazza:2009zz} as a means to experimentally test the LL equation via the resulting Thomson scattering spectra.  For a head on collision (i.e. $\theta=180^\circ$) between an electron and a short plane wave laser pulse the authors derive an expression for the onset of reflection.  This is
\begin{eqnarray}
R\gtrsim \frac{4\gamma_0^2-a_0^2}{2a_0^2}>0,\quad a_0\gg 1,\label{DPreflection}
\end{eqnarray}
which, below the radiation dominated regime R, reduces to $2\gamma_0>a_0$.\footnote{Further light may be shed on this by considering the kinematics in terms  of a laser induced mass shift of the electron.  If one averages the electron momentum over a laser cycle, one finds that the electron can be considered to have undergone a mass shift $m^2\longrightarrow m^2(1+a_0^2)$.  Then one finds that when $a_0\approx 2\gamma_0$ the laboratory frame can be interpreted as an intensity-dependent centre-of-mass frame.  See \cite{Harvey:2009ry} for a fuller discussion.}  This is broadly consistent with our results, bearing in mind that we are not considering a directly head on collision.

To investigate this phenomena more fully we will fix our intensities at $a_0=180$ and 250 and (for both the undamped and the damped solutions) determine numerically the size of the longitudinal shift of the electron as a function of $\gamma_0$.  We define our longitudinal displacement parameter $D$ to be the longitudinal shift that the electron undergoes before it reaches a distance $\vert x \vert =20\mu$m from the laser axis, compared to where it would have been if no field was present.  (It should be noted that our displacement parameter $D$ measures a quantity that is not directly analogous to the `reflection' referred to in \cite{DiPiazza:2009zz}.  This is because our $D$ parameter measures the field induced change in longitudinal position of the electron at a \text{distance} from the beam axis.  Thus this measure registers angular \textit{deflections} as well as direct reflections, which is precisely what a detector would measure if it were placed alongside the beam.)  The displacement parameter is shown in the left hand plots of Figure \ref{fig:pw_disp} and the net energy gain ($\gamma_f -\gamma_0$) on the right.

At low energies the electrons are strongly displaced/reflected, but experience only a minimal change in energy.  A useful criteria for demarking this region is found to be $\gamma_0$ such that $R=0.01$, i.e. $1\%$ of the radiation dominated regime.  As we increase $\gamma_0$ above this value we find that the radiation damping begins to play a role in the dynamics, with the damped electron being subject to a significantly higher displacement than its undamped counterpart.  In this regime the damped electron loses a large proportion of its initial energy, while the undamped one remains unaffected.  Beyond this, in the reflection regime $2\gamma_0>a_0$, it is only the damped electron that experiences a displacement.   This, together with our results in Figure \ref{fig:FigurePW}, indicate that we have entered a regime of \textit{radiation reaction induced} electron capture.

\subsection{Gaussian beam dynamics}\label{gaussiansection}
\begin{figure}[!h]
\includegraphics[scale=1.0,clip=true,viewport=90 230 600 580]{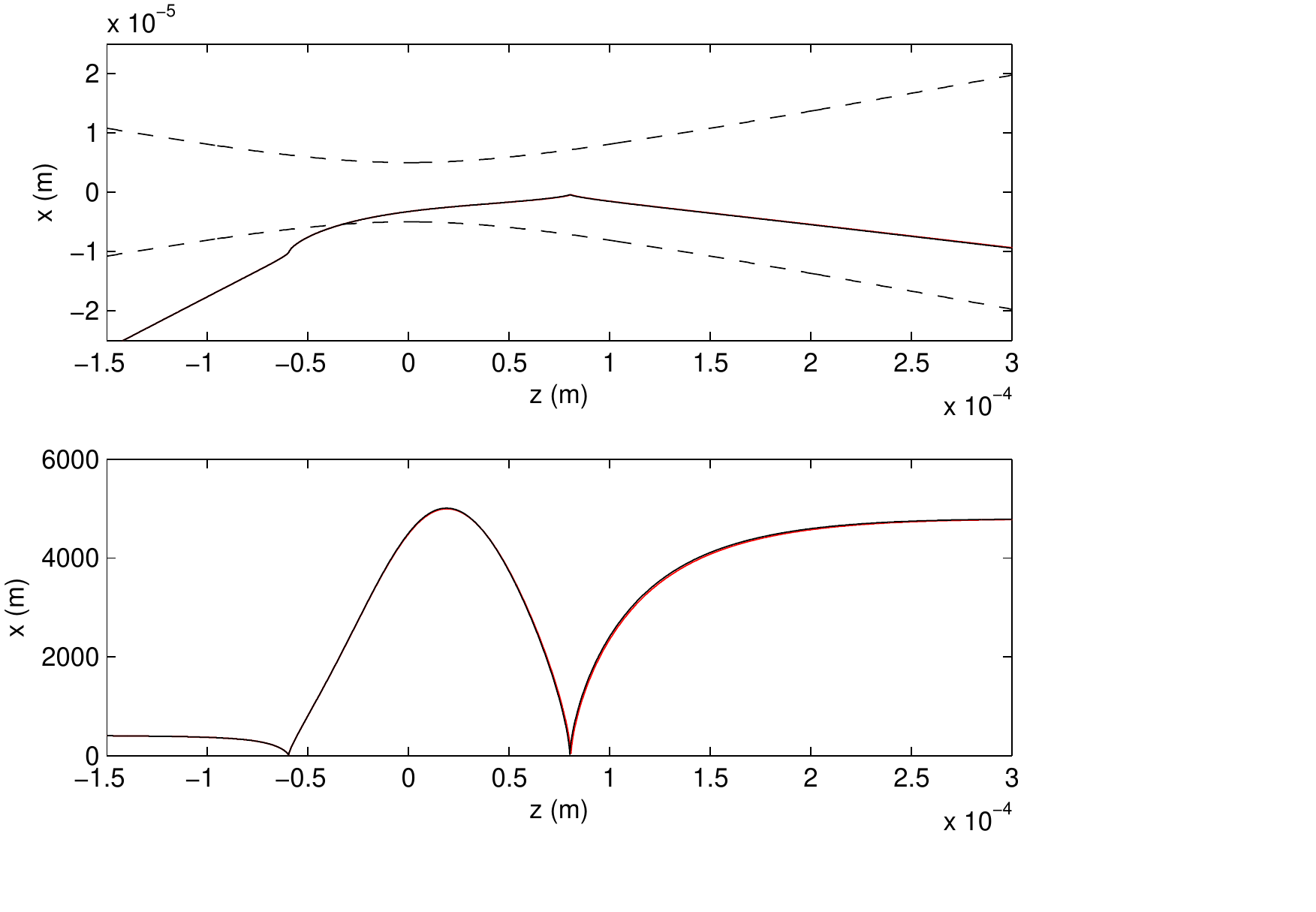}
\caption{\label{fig:FigureB} Near same direction collision ($\theta=10$ degrees) between an electron and a pulsed Gaussian beam.  Parameters are  $a_0=600$, $\gamma =400$, $\eta_0 =20$.  Black lines (solid): undamped solutions, red lines: damped solutions.  Black lines (dashed): beam waist. 
}
\end{figure}
Having throughly considered the plane wave dynamics, we now turn our attention to the more realistic pulsed Gaussian beam model of the laser.  We begin by looking at the case of a near same direction collision.  As stated earlier, in this case an electron in its rest frame will see a laser of lower intensity than if it were counter propagating with the beam.  The advantage of this set up is that it allows for a CAS scenario where the electron will get captured and carried along with the pulse.  Thus radiation damping effects, although somewhat smaller, may accumulate over a longer time period.  Choosing our parameters carefully so that CAS takes place, one finds that, even though the energy gains of the electron are large (see Figure \ref{fig:FigureB}), the effect of radiation damping is negligible.  This is true even for the very highest parameter values allowed in the classical regime and is consistent with the results of Mao \textit{et al} \cite{Mao:2010}

\begin{figure}[!h]
\includegraphics[scale=0.8,clip=true,viewport=10 230 600 580]{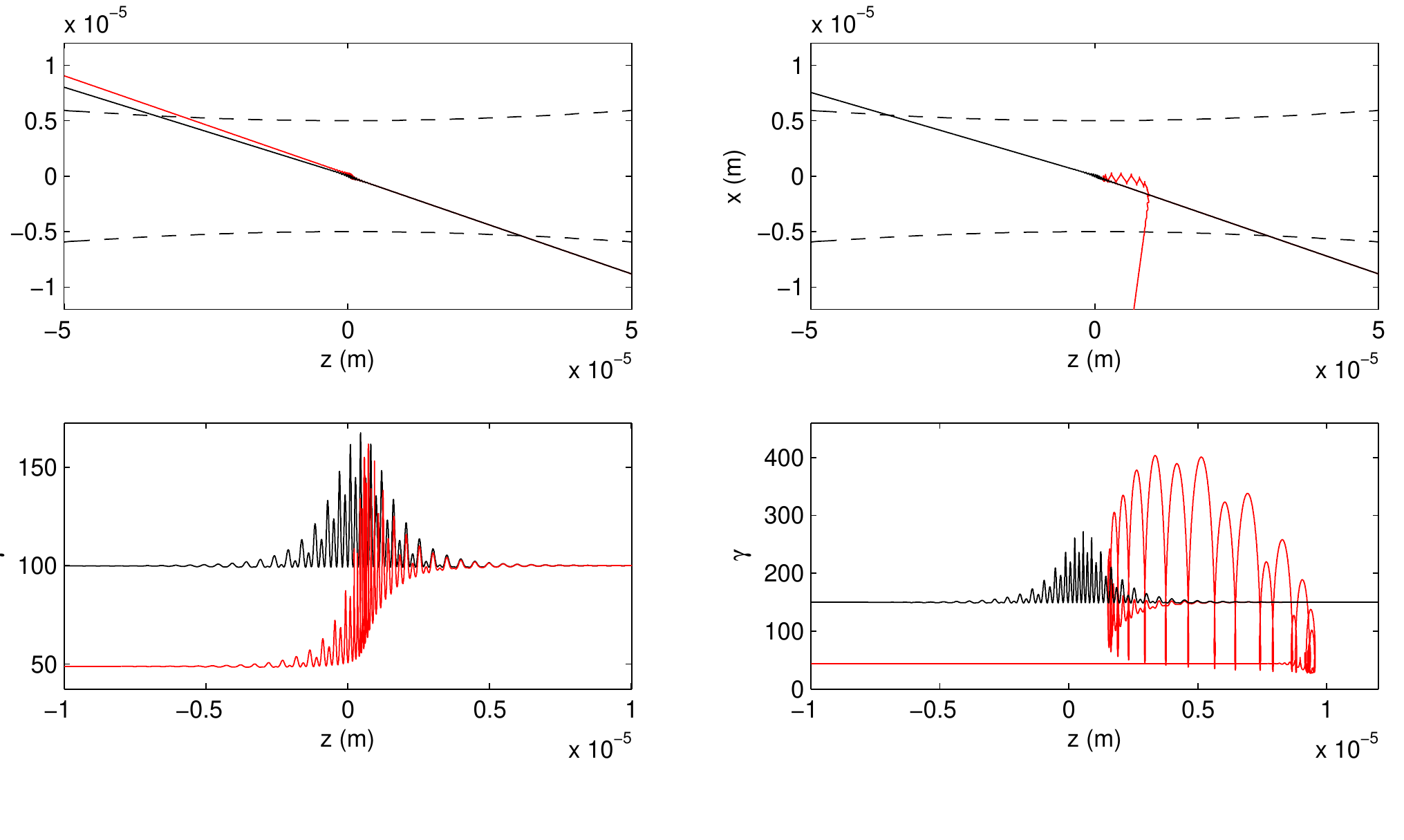}
\caption{\label{fig:FigureA} Near head on collision ($\theta=170$ degrees).  Left plots: $a_0=150$, $\gamma_0 =100$; right plots:  $a_0=250$, $\gamma_0 =150$; $\eta_0 =20$ for both cases.  Top plots show trajectories and bottom plots show corresponding electron energies.  Black lines (solid): undamped solutions, red lines: damped solutions, black lines (dashed): beam waist $w$.}
\end{figure}

Therefore we now turn our focus to the case of a near head on collision.  In Figure \ref{fig:FigureA} we show the electron trajectories and energies for the same parameter values as the plane wave case in Figure \ref{fig:FigurePW}.  For the more modest parameter values in the left hand plots ($a_0=150$, $\gamma_0=100$) we find that the behaviour of both the damped and undamped electrons is very similar to the plane wave case, with little reflection and radiation damping causing an energy loss of roughly 50\%.  At the higher parameter values in the right hand plots ($a_0=250$, $\gamma_0=150$ -- which are again towards the higher end of what is allowed in the classical regime) the dynamics have altered somewhat from the plane wave case.  Although the damped electron is reflected by the beam, it leaves it in a different direction.  Nevertheless the electron $\gamma$-factor has a very similar behaviour to the plane wave case.

\begin{figure}[ht!]
\includegraphics[scale=0.8,clip=true,viewport=10 215 600 650]{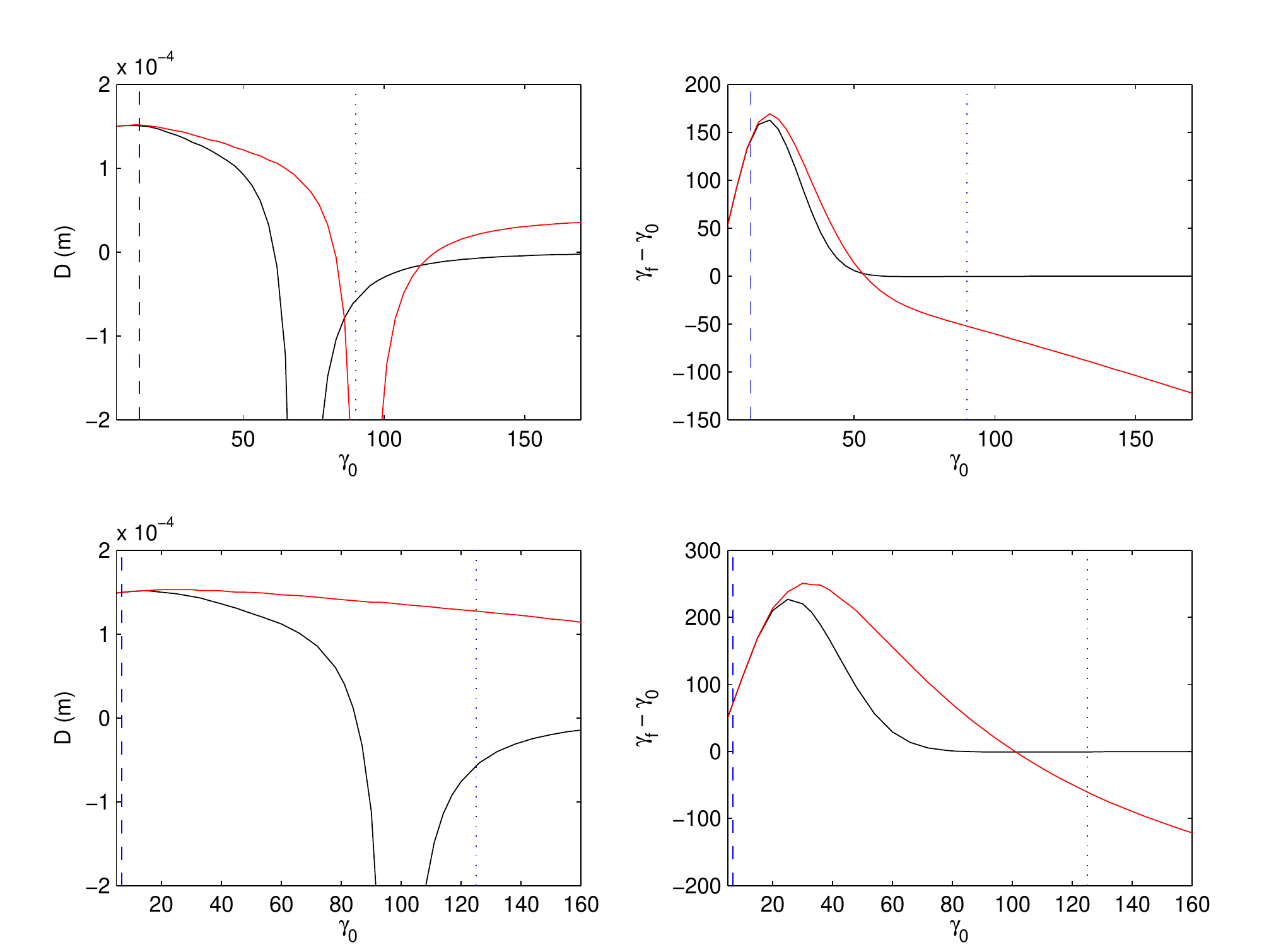}
\caption{\label{fig:pulse_disp} Left plots: longitudinal displacement $D$ of the electron as a function of $\gamma_0$ for a Gaussian pulse, near same direction collision ($\theta=170$ degrees).  Right plots: Net energy change $\gamma_f-\gamma_0$ as a function of $\gamma_0$. Top plots have parameters $a_0=180$, $\eta_0 =20$ and bottom plots $a_0=250$, $\eta_0 =20$ .  Black lines:  undamped solutions, red lines: damped solutions, blue dotted lines indicate $1\%$ of $R$, blue dashed lines: $2\gamma_0=a_0$.
}
\end{figure}
\begin{figure}
\includegraphics[scale=0.7,clip=true,viewport=90 240 600 570]{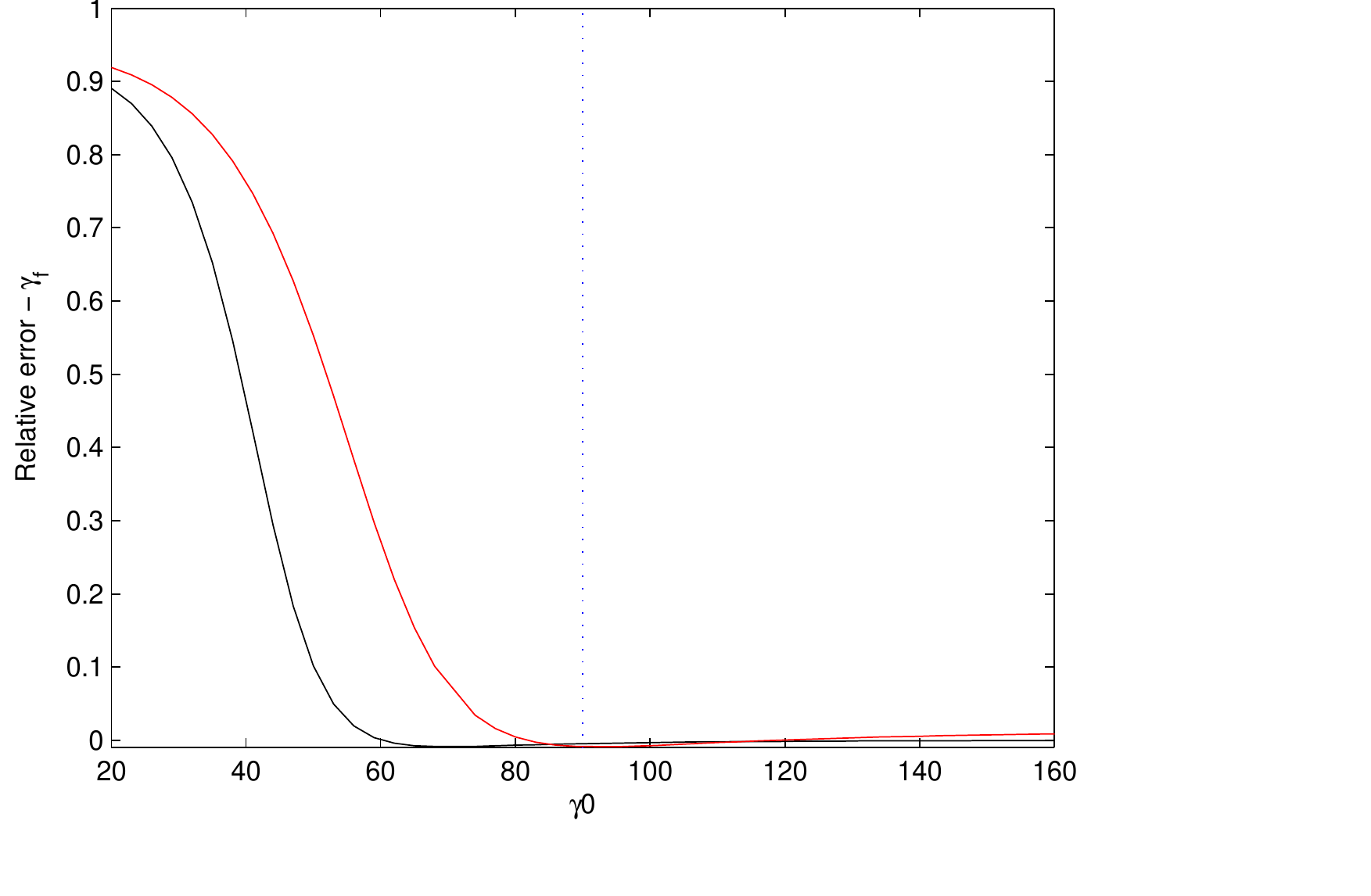}
\caption{\label{fig:errors_pw_v_pulse} Relative error of the final electron energy $\gamma_f$ calculated using the plane wave approximation and the full Gaussian beam expression, $(\gamma_\textrm{Gauss}-\gamma_\textrm{PW})/\gamma_\textrm{Gauss}$.  Parameters are $a_0=180$, $\eta_0=20$.  Black lines: undamped solution, red lines: damped solution, blue dotted line: $2\gamma_0=a_0$.
}
\end{figure}

As with our plane wave analysis, it is instructive to fix the laser intensity and consider the longitudinal displacement and final electron energy as a function of $\gamma_0$.  The results of this analysis are shown in Figure \ref{fig:pulse_disp}.  Once again there are three distinct regimes, although the criteria we used to demarcate the first of these regions in the plane wave case is now only suitable as a rough guide.  Generally, we find that the radiation damping effects kick in later than for the plane wave case.  This can be seen from the plots by observing that the damped and undamped lines are still together beyond the $1\% R$ limit (which is where they separated in our plane wave examples).  That the radiation damping effects occur later for the Gaussian pulse field is to be expected, since the field intensity decays more quickly away from the focal point than in the plane wave model.  Observe that in the case of low $\gamma_0$ both the damped and undamped electrons experience a net energy gain, whereas in the plane wave case it was only the damped electron that underwent an energy change.  The marker $2\gamma_0 =a_0$ still serves as a good indicator as to when the undamped electron can pass through the beam without being displaced.  Therefore, once again, in the regime $2\gamma_0 >a_0$ we have a situation where the damped electron is displaced but the undamped one is not.  Examining the trajectories (e.g. Figure \ref{fig:FigureA}) we see that we are in a regime of radiation reaction induced electron capture.  Comparing Figures \ref{fig:pulse_disp} and \ref{fig:pw_disp} indicates that in this regime the displacements and energy gains are well described by the plane wave model.  To investigate further, in Figure \ref{fig:errors_pw_v_pulse} we plot the relative error of the plane wave approximation to the final energy $(\gamma_\textrm{Gauss}-\gamma_\textrm{PW})/\gamma_\textrm{Gauss}$.  Beyond $2\gamma_0=a_0$ the plane wave model gives us an extremely good measure of the net energy change, even though the actual dynamics are a little different.  That the two models converge can be understood by recalling that the high intensity limit is the same as the low frequency limit.  In the low frequency limit both the plane wave and Gaussian beam models behave as constant crossed fields.

\begin{figure}[!h]
\includegraphics[scale=0.7,clip=true,viewport=90 230 600 580]{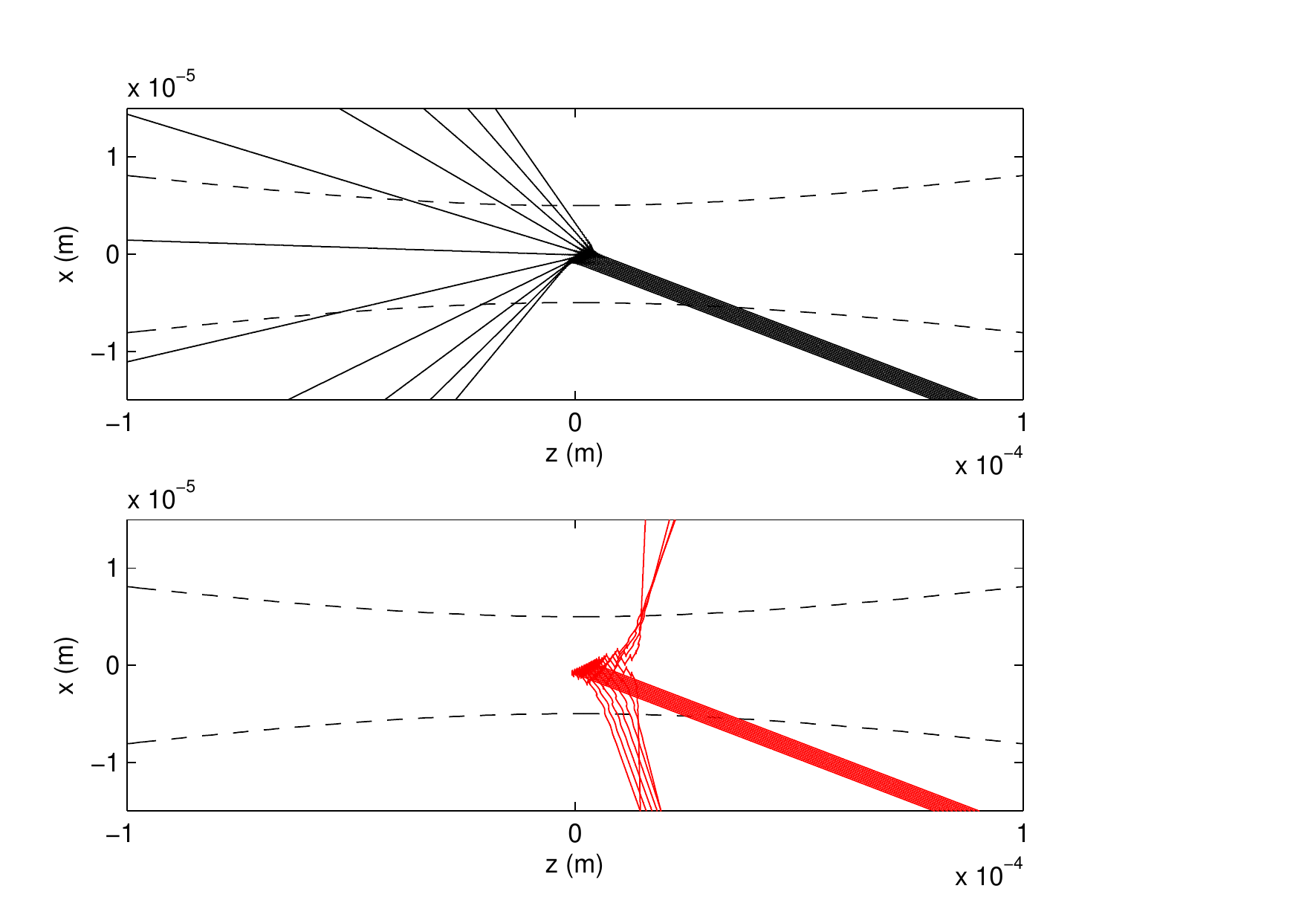}
\caption{\label{fig:multiple1} Near head on collision ($\theta=170$ degrees).  Parameters are  $a_0=250$, $\gamma =100$, $\eta_0 =20$.  Black lines (solid): undamped solutions, red lines: damped solutions, black lines (dashed): beam waist $w$.
}
\end{figure}
\begin{figure}[!h]
\includegraphics[scale=0.7,clip=true,viewport=90 230 600 580]{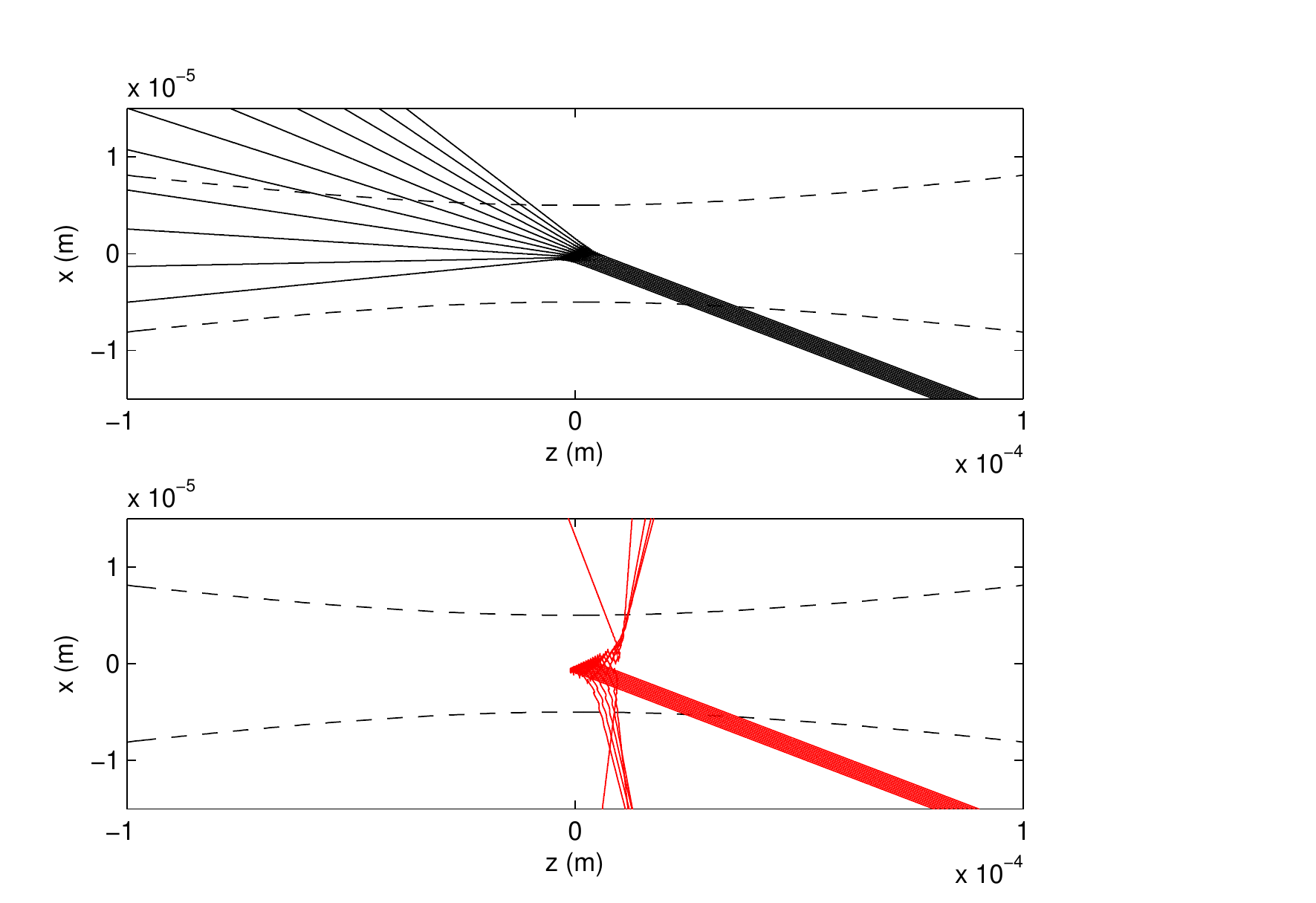}
\caption{\label{fig:multiple2} Near head on collision ($\theta=170$ degrees).  Parameters are  $a_0=250$, $\gamma =150$, $\eta_0 =20$.  Black lines (solid): undamped solutions, red lines: damped solutions, black lines (dashed): beam waist $w$.
}
\end{figure}

\begin{figure}
\includegraphics[scale=0.8,clip=true,viewport=90 230 600 580]{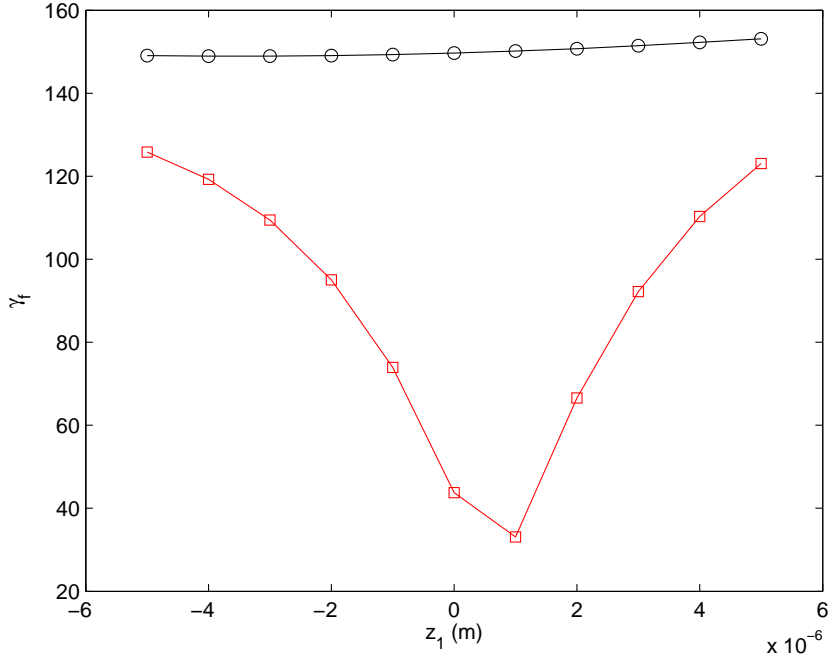}
\caption{\label{fig:multiple_final_gamma} Final gamma as a function of $z_1$.  Near head on collision ($\theta=170$ degrees).  Parameters are  $a_0=250$, $\gamma =150$, $\eta_0 =20$.  Black lines/circles: undamped solutions, red lines/squares: damped solutions.
}
\end{figure}

In a realistic situation it will not be possible to ensure that the electrons will all be aiming precisely for the centre of the beam focus.  For our results to be interesting experimentally, we need to consider their sensitivity to the spatial size of the electron beam.  Therefore we conduct some simulations where small bunches of electrons interact with the beam.  A typical source of electrons in experiments such as these is from a linac.  In our simulations we will model the electron beam as being $10\mu$m in diameter, which is typical of facilities such as the ELBE accelerator at the Forschungszentrum Dresden-Rossendorf in Germany \cite{Arnold:2007}.  We will assume the beam to be of a high quality with negligible energy spread (e.g. $\Delta\gamma/\gamma_0 =10^{-3}$ is feasible with the aforementioned facility \cite{Heinzl:2009nd}) and that the concentration of electrons is dilute enough for Coulomb repulsion between the electrons to be neglected.  The bunches will be centered at the beam focus, extending to 5$\mu$m either side of it (i.e. $z_1 =-5\mu$m$\ldots 5\mu$m), with all the electrons travelling parallel to one another.  The resulting particle trajectories with and without damping are shown in Figures \ref{fig:multiple1} and \ref{fig:multiple2}.  In Figure \ref{fig:multiple1} we choose our parameter values ($\gamma_0=100$, $a_0=250$) from the region $[R>1\%, 2\gamma_0<a_0]$, i.e. the regime where both the damped and undamped electrons experience displacement (see Figure \ref{fig:pulse_disp}).  One sees that the damped and undamped dynamics are very different to one another.  The scattering direction of the undamped electrons is much more sensitive to $z_1$ than for the damped electrons.  The behaviour of the damped electrons is more consistent, with all of them being reflected, although the exit direction flips as we move slightly in front of the laser focus.  For the second set of parameter values (Figure \ref{fig:multiple2}, $\gamma_0=150$, $a_0=250$) we are in the radiation reaction induced capture regime ($2\gamma_0>a_0$) where the electrons have sufficient energy so that the undamped solutions exhibit only minimal displacement, but radiation damping is strong enough to capture the damped electrons.  In this regime the undamped electrons are well culminated, exhibiting much less deflection than in the previous example and, once again, all of the damped electrons are captured by the beam.  We find that, although there is still a flip in the direction of the ejected electrons, the essential phenomena of radiation reaction induced capture is stable with respect to electron beam width effects.  The net energy change of the damped electrons is a little more sensitive to $z_1$, however, as can be seen by considering the final $\gamma$-factors of these electrons (Figure \ref{fig:multiple_final_gamma}).  

\subsection{XFELs}
Let us now briefly consider how the situation changes if we swap our optical laser field for an XFEL beam.  The properties of XFEL beams are somewhat different to the optical fields we have been considering \cite{XFEL}, and this will reflect itself in the subsequent dynamics.  As well as the obvious shorter wavelength, the main differences are the weaker focusing of the beam and the much higher number of cycles in the pulse.  The weaker focusing means that the beam waist is many laser wavelengths in diameter.  A typical 100fs pulse with wavelength $\lambda=1$nm will contain of order $10^5$ cycles.  Thus the XFEL pulse behaves almost as if it were a continuous beam with stationary focus (i.e. $g\sim 1$).  This means that the electron will spend a longer time in a weaker part of the pulse before reaching the focus, thus needing a comparatively higher initial energy in order to probe the most intense part of the beam.  The upshot of this is that the spatial effects of the beam will play a bigger role, which may mask some of the radiation damping effects.

\begin{figure}[ht!]
\includegraphics[scale=0.8,clip=true,viewport=10 215 600 650]{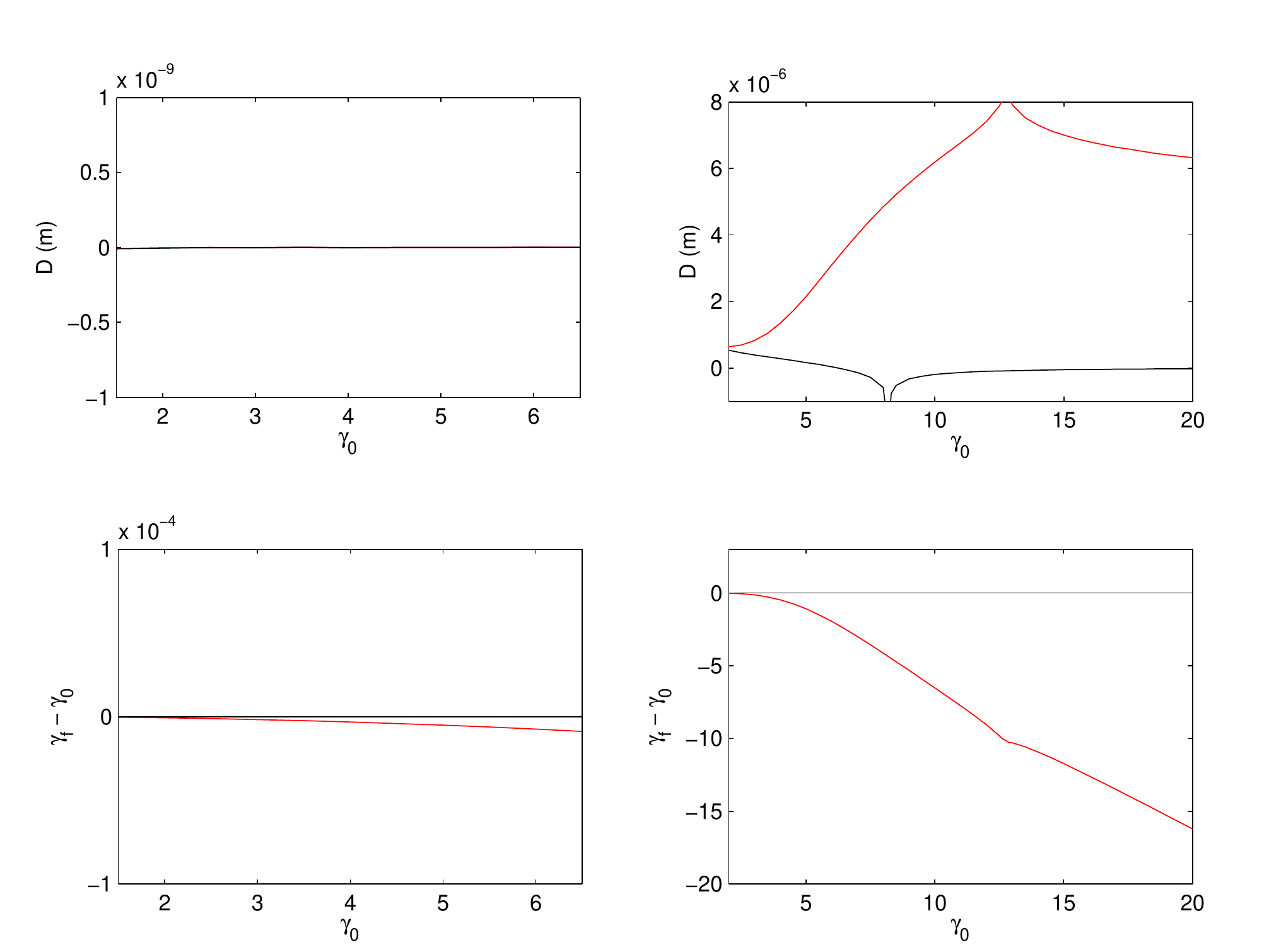}
\caption{\label{fig:XFEL} Study of the dynamics in an XFEL beam.  Parameters are $w_0=0.05\mu$m, $\lambda=1$nm, $\theta=170$ degrees and $\eta_0=1.89\times 10^5$, implying a pulse duration of $\sim100$fs.  Left plots: $a_0=0.002$, right plots: $a_0=2$.  Top plots: longitudinal displacement $D$ of the electron as a function of $\gamma_0$ (Note that the values are calculated for a point $2\mu$m from the beam axis, rather than $20\mu$m as they were in the optical examples.).  Bottom plots: net energy change $\gamma_f-\gamma_0$ as a function of $\gamma_0$.  Black lines: undamped solutions, red lines: damped solutions.}
\end{figure}

In Figure \ref{fig:XFEL} we show the results of a numerical study of the dynamics of electrons in typical XFEL fields.  The left hand plots show the electron displacement and net energy change for the case of a 1nm wavelength field of intensity $a_0=0.002$.  This intensity corresponds to $10^{19}$W/cm$^2$ and is likely to be the highest that can be achieved with the newest generation of facilities, such as the European XFEL \cite{XFEL}.  It is clear from the plots that, even though we are far into the regime $2\gamma_0>a_0$, the intensity is just not high enough for radiation damping to be significant.  We could of course increase the electron energy but, since the electron energy would then be far greater than the laser intensity, the interaction time would be very short.  Also, and perhaps more importantly, the derivation of the LL equation is not valid for higher $\gamma_0$ with these parameter values (see (\ref{llconstraint})) and so we would have to find an alternative approach to describing the dynamics.  Instead we consider the dynamics with a larger $a_0$.  The right hand plots of Figure \ref{fig:XFEL} show the same quantities but with $a_0=2$.  For these values the effects of radiation damping are once again significant, resulting in an increased displacement and net energy loss to the electrons.  However, upon examining the trajectories we find that the difference in displacement is caused by a \textit{deflection} of the damped electron rather than a \textit{reflection}.  Therefore we don't have the same phenomena of radiation reaction induced electron capture as we did in the optical case.

\section{Conclusions and Summary}
We have analysed the impact of radiation damping on the dynamics of electrons in high intensity pulsed laser fields.  We began by considering an idealised pulsed plane wave model of the laser field in a `near head on' ($\theta=170$ degrees) collision with a counter propagating electron.  For relatively low parameter values ($\gamma_0=100$, $a_0=150$) we confirmed that the radiation damping causes an energy loss to the electron (in this case $\sim 50$\%) but has only a minimal impact on its trajectory.  At high intensities it was found that the damped electron can become reflected by the beam and we saw evidence of radiation damping induced electron capture.  Performing a more detailed analysis, we found that the dynamics fall neatly into three regimes.  For low $\gamma_0$, such that $R<1\%$, the electrons are strongly deflected by the pulse but the effects of radiation damping are negligible.  If we increase $\gamma_0$ such that we are in the range $[R>1\%, 2\gamma_0<a_0]$, then damping effects come into play.  Both damped and undamped electrons undergo displacement, but the damped electron experiences a net energy loss.  In the very high energy regime, $2\gamma_0>a_0$, the damped electron is displaced but the undamped one is not.  We then moved to the more realistic paraxial pulsed Gaussian beam model of the laser field.  This allowed us to take into account the spatial as well as the temporal effects of the beam.  Taking our beam to be focused to a waist of $w_0=5\mu$m, we considered two different intensities -- $a_0=180$ and 250 -- and studied how the subsequent dynamics change as a function of the initial electron energy $\gamma_0$.  We found once again that the dynamics fall into three regimes.  For low $\gamma_0$ the radiation damping effects are again negligible.  However, damping effects start to become important at higher energies than for the plane wave examples, with $R=1\%$ no longer being a reliable indicator of the onset of the damping regime.  Also unlike the plane wave case, for low $\gamma_0$ both the damped and undamped electrons in the Gaussian pulse experience a net energy gain.  In the intermediate region $[\gamma_0\gg 1, 2\gamma_0 <a_0]$ both the damped and undamped electrons were found to be displaced/deflected by the beam.  The condition $2\gamma_0=a_0$ once again provided a good criteria for when the undamped electron is no longer deflected.  In the high energy regime $2\gamma_0>a_0$ it was found that only the damped electron gets displaced.  Thus one can identify this regime as one of radiation damping induced electron capture.  Another feature of this regime is that the electron dynamics are qualitatively the same for the Gaussian beam as for the plane wave, with the plane wave model providing an extremely good approximation to the net energy change.  Also, giving consideration to the spatial effects of the electron beam, we found the dynamics to be reasonably stable with respect to changes in the electron position.  Finally, radiation damping in XFEL type fields was briefly considered.  It was found that the relatively low intensities expected in the current generation of facilities will limit any damping effects.

Thus we have analysed the effects of radiation damping in a realistic model of a laser field.  What remains is for our results to  be tested experimentally.  An experimental test of our radiation reaction induced electron capture scenario, in particular, will allow an evaluation of the applicability of the classical theory in the context of such high intensity laser fields.

\acknowledgements
The authors thank Tom Heinzl, Anton Ilderton, Kurt Langfeld, Hartmut Ruhl and Marija Vranic for fruitful discussions.  C. H. is greatly indebted to Amol Holkundkar for help with coding issues.  C. H. was supported by the Swedish Research Council Contract \# 2007-4422 and the European Research Council Contract \# 204059-QPQV.


\end{document}